\documentclass[aps,prl,twocolumn,groupedaddress,superscriptaddress,amsfonts,amssymb,amsmath,citeautoscript,a4paper]{revtex4-1}

\usepackage[utf8]{inputenc}
\usepackage[english]{babel}
\usepackage{microtype}
\usepackage{txfonts}
\usepackage{txfontsb}

\usepackage{bm}
\usepackage{xcolor}
\usepackage{graphicx}
\usepackage{float}
\usepackage{xspace}

\usepackage{enumerate}
\usepackage[inline]{enumitem}

\usepackage{siunitx}
\DeclareSIUnit[number-unit-product=]\percent{\char`\%} 

\usepackage[pdftex]{hyperref}
\hypersetup{colorlinks,
            linkcolor={blue!75!black!80!yellow},
            citecolor={blue!75!black!80!yellow},
            urlcolor={blue!75!black!80!yellow},
            pdfstartview=FitH}
            
\usepackage[eulergreek]{sansmath}
\makeatletter
\renewcommand\@make@capt@title[2]{%
    \@ifx@empty\float@link{\@firstofone}{\expandafter\href\expandafter{\float@link}}%
    \sisetup{math-sf=\textsf}%
    \sansmath\sffamily\textbf{#1\@caption@fignum@sep}#2
}%

\makeatother

\usepackage{etoolbox}

\newcommand{\ie}{i.e.\@\xspace}  

\newcommand{\eg}{e.g.\@\xspace}

\newcommand{\ep}{\epsilon}
\newcommand{\pll}{\parallel}
\newcommand{\EF}{E_\text{F}}
\newcommand{\e}{\text{e}}
\newcommand{\iu}{\text{i}}
\let\Re\relax\DeclareMathOperator{\Re}{Re}
\let\Im\relax\DeclareMathOperator{\Im}{Im}

\newcommand{\appropto}{\mathrel{\vcenter{
			\offinterlineskip\halign{\hfil$##$\cr
				\propto\cr\noalign{\kern.2pt}\sim\cr\noalign{\kern-2.5pt}}}}}

\newcommand{\SDUaffil}{\footnotesize Center for Nano Optics, University of Southern Denmark, 5230 Odense M, Denmark}
\newcommand{\MITaffil}{\footnotesize Department of Physics, Massachusetts Institute of Technology, Cambridge, MA 02139, USA}
\newcommand{\CNGaffil}{\footnotesize Center for Nanostructured Graphene, Technical University of Denmark, 2800 Kgs. Lyngby, Denmark}
\newcommand{\DIASaffil}{\footnotesize Danish Institute for Advanced Study, University of Southern Denmark, 5230 Odense M, Denmark}
\newcommand{\DTUPHYSaffil}{\footnotesize Department of Physics, Technical University of Denmark, 2800 Kgs. Lyngby, Denmark}
\newcommand{\UMaffil}{\footnotesize Department of Physics and Center of Physics, University of Minho, 4710-057 Braga, Portugal}
\newcommand{\INLaffil}{\footnotesize International Nanotechnology Laboratory, Av. Mestre Jos\'{e} Veiga, 4710-330 Braga, Portugal}
\newcommand{\ICFOaffil}{\footnotesize ICFO -- Institut de Ci\`{e}ncies Fot\`{o}niques, The Barcelona Institute of Science and Technology, 08860 Castelldefels (Barcelona), Spain}
\newcommand{\ICREAaffil}{\footnotesize ICREA -- Instituci{\'o} Catalana de Recera i Estudis Avan\c{c}ats, 08010 Barcelona, Spain}

\begin{document}
\title{Quantum Surface-Response of Metals Revealed by Acoustic Graphene Plasmons}

\author{P.~A.~D.~Gon\c{c}alves}
\email{pa@mci.sdu.dk}
\affiliation{\MITaffil} \affiliation{\SDUaffil}  

\author{Thomas~Christensen}
\affiliation{\MITaffil}

\author{Nuno~M.~R.~Peres}
\affiliation{\UMaffil} \affiliation{\INLaffil}

\author{Antti-Pekka~Jauho}
\affiliation{\CNGaffil} \affiliation{\DTUPHYSaffil}

\author{Itai~Epstein}
\affiliation{\ICFOaffil}
\affiliation{\footnotesize Department of Physical Electronics, School of Electrical Engineering, Tel Aviv University, Tel Aviv 6997801, Israel}

\author{Frank~H.~L.~Koppens}
\affiliation{\ICFOaffil}
\affiliation{\ICREAaffil}

\author{Marin~Solja\v{c}i\'{c}}
\affiliation{\MITaffil}

\author{N.~Asger~Mortensen}
\email{asger@mailaps.org}
\affiliation{\SDUaffil} \affiliation{\CNGaffil} \affiliation{\DIASaffil}


\begin{abstract}
 A quantitative understanding of the electromagnetic response of materials is essential for the precise engineering of maximal, versatile, and controllable light--matter interactions. Material surfaces, in particular, are prominent platforms for enhancing electromagnetic interactions and for tailoring chemical processes.
 However, at the deep nanoscale, the electromagnetic response of electron systems is significantly impacted by quantum surface-response at material interfaces, which is  challenging to probe using standard optical techniques.
 Here, we show how ultraconfined acoustic graphene plasmons in graphene--dielectric--metal structures can be used to probe the quantum surface-response functions of nearby metals, here encoded through the so-called Feibelman $d$-parameters. 
 Based on our theoretical formalism, we introduce a concrete proposal for experimentally inferring the low-frequency quantum response of metals from quantum shifts of the acoustic graphene plasmons dispersion, 
 and demonstrate that the high field confinement of acoustic graphene plasmons can resolve intrinsically quantum mechanical electronic length-scales with subnanometer resolution. 
 Our findings reveal a promising scheme to probe the quantum response of metals, and further suggest the utilization of acoustic graphene plasmons as plasmon rulers with \r{a}ngstr\"{o}m-scale accuracy. 
\end{abstract}

\maketitle


\section{Introduction}

Light is a prominent tool to probe the properties of materials and their electronic structure, as evidenced by the widespread use of light-based spectroscopies across the physical sciences~\cite{OptSpecBook,ModSpecBook}.
Among these tools, far-field optical techniques are particularly prevalent, but are constrained by the diffraction limit and the mismatch between optical and electronic length scales to probe the response of materials only at large length scales (or, equivalently, at small momenta).
Plasmon polaritons---hybrid excitations of light and free carriers---provide a mean to overcome these constraints through their ability to confine electromagnetic radiation to the nanoscale~\cite{Gramotnev:2010}.

Graphene, in particular, supports gate-tunable plasmons characterized by an unprecedentedly strong confinement of light~\cite{GoncalvesPeres,Koppens:2011,FJGAbajo:2014}.
When placed near a metal, graphene plasmons (GPs) are strongly screened and acquire a nearly-linear (acoustic-like) dispersion~\cite{Alonso:2017,Lundeberg:2017,Goncalves_SpringerTheses,Epstein:2020} (contrasting with the square-root-type dispersion of conventional GPs). Crucially, such acoustic graphene plasmons (AGPs) in graphene--dielectric--metal (GDM) structures have been shown to exhibit even higher field confinement than conventional GPs with the same frequency, effectively squeezing light into the few-nanometer regime~\cite{Lundeberg:2017,Goncalves_SpringerTheses,Iranzo:2018,Epstein:2020}.
Recently, using scanning near-field optical microscopy, these features were exploited to experimentally measure the conductivity of graphene, $\sigma(q,\omega)$, across its frequency ($\omega$) and momentum ($q$) dependence simultaneously~\cite{Lundeberg:2017}.
The observation of momentum-dependence implies a nonlocal response (\ie, response contributions at position $\mathbf{r}$ from perturbations at $\mathbf{r}'$), whose origin is inherently quantum mechanical.
Incidentally, traditional optical spectroscopic tools cannot resolve nonlocal response in extended systems due to the intrinsically small momenta ${k_0 \equiv \omega/c}$ carried by far-field photons. 
Acoustic graphene plasmons, on the other hand, can carry large momenta---up to a significant fraction of the electronic Fermi momentum $k_\text{F}$ and with group velocities asymptotically approaching the electron's Fermi velocity $v_\text{F}$---and so can facilitate explorations of nonlocal (\ie, $q$-dependent) response not only in graphene itself, but also, as we detail in this Article, in nearby materials.
So far, however, only aspects related to the quantum response of graphene have been addressed~\cite{Lundeberg:2017}, leaving any quantum nonlocal aspects of the adjacent metal's response unattended, despite their potentially substantial impact at nanometric graphene--metal separations~\cite{Yan:2015,Christensen:2017,Dias:2018,Yang:2019,Goncalves:2020}.

Here, we present a theoretical framework that simultaneously incorporates quantum nonlocal effects in the response of both the graphene and of the metal substrate for AGPs in GDM heterostructures.
Further, our approach establishes a concrete proposal for experimentally measuring the low-frequency nonlocal electrodynamic response of metals. 
Our model treats graphene at the level of the nonlocal random-phase approximation (RPA)~\cite{Wunsch:2006,Hwang:2007,Jablan:2009,GoncalvesPeres,Goncalves_SpringerTheses} 
and describes the quantum aspects of the metal's response---including nonlocality, electronic spill-out/spill-in, and surface-enabled Landau damping---using a set of microscopic surface-response functions known as the Feibelman $d$-parameters~\cite{Feibelman:1982, LiebschBook, Yan:2015, Christensen:2017, Yang:2019, Goncalves:2020}. 
These parameters, $d_\perp$ and $d_\pll$, measure the frequency-dependent centroids of the induced charge density and of the normal derivative of the tangential current density, respectively (Supplementary Note~1).
Using a combination of numerics and perturbation theory, we show that the AGPs are spectrally shifted by the quantum surface-response of the metal: toward the red for $\Re d_\perp > 0$ (associated with electronic spill-out of the induced charge density) and toward the blue for $\Re d_\perp < 0$ (signaling an inward shift, or ``spill-in'').
Interestingly, these shifts are not accompanied by a commensurately large quantum broadening nor by a reduction of the AGP's quality factor, 
thereby providing the theoretical support explaining recent experimental observations~\cite{Iranzo:2018}.
Finally, we discuss how state-of-the-art measurements of AGPs could be leveraged to map out the low-frequency quantum nonlocal surface-response of metals experimentally. Our findings have significant implications for our ability to optimize photonic designs that interface far- and mid-infrared optical excitations---such as AGPs---with metals all the way down to the nanoscale, with pursuant applications in, \eg, ultracompact nanophotonic devices, nanometrology, and in the surface sciences more broadly.

\begin{figure}[t]
 \centering
  \includegraphics[width=1.0\columnwidth]{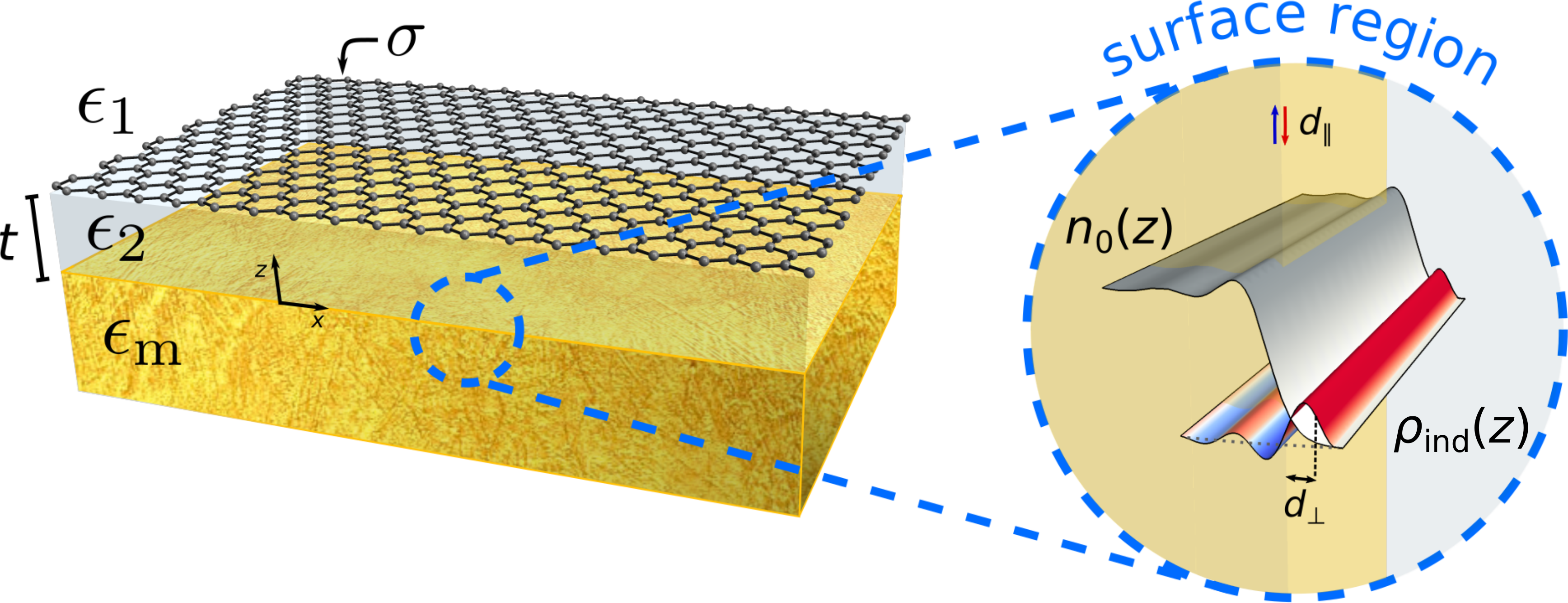}
  \caption{\textbf{Schematics of a dielectric--graphene--dielectric--metal (GDM) heterostructure.}
  The graphene--metal separation, $t$, is controlled by the thickness of the dielectric ($\ep_2$) spacer.
  The close-up (near the metal--spacer interface) shows a pictorial representation of the surface-response functions $d_\perp$ and $d_\pll$ along with the related the microscopic quantities characterizing the metal surface, namely the equilibrium electronic density, $n_0(z)$, and the induced charge density, $\rho_{\text{ind}}(z)$.
  }\label{fig:GDM_scheme_Feib}
\end{figure}
%


\section{Results}

\noindent
\textbf{Theory.} We consider a GDM heterostructure (see Fig.~\ref{fig:GDM_scheme_Feib}) composed of a graphene sheet with a surface conductivity $\sigma \equiv \sigma(q,\omega)$ separated from a metal substrate by a thin dielectric slab of thickness $t$ and relative permittivity $\ep_\text{2} \equiv \ep_\text{2}(\omega)$; finally, the device is covered by a superstrate of relative permittivity $\ep_\text{1} \equiv \ep_\text{1}(\omega)$.
While the metal substrate may, in principle, be represented by a nonlocal and spatially non-uniform (near the interface) dielectric function, here we abstract its contributions into two parts: a bulk, local contribution via $\ep_{\text{m}} \equiv \ep_{\text{m}}(\omega) = \ep_\infty(\omega) - \omega_{\text{p}}^2/(\omega^2 + \iu \omega \gamma_{\text{m}})$, and a surface, quantum contribution included through the $d$-parameters. 
These parameters are quantum-mechanical surface-response functions, defined by the first moments of the microscopic induced charge ($d_\perp$) and of the normal derivative of the tangential current ($d_\pll$); see Fig.~\ref{fig:GDM_scheme_Feib} (Supplementary Note~1 gives a concise introduction).
They allow the leading-order corrections to classicality to be conveniently incorporated via a surface dipole density ($\propto d_\perp$) and a surface current density ($\propto d_\pll$)~\cite{Yang:2019,Goncalves:2020,Goncalves_SpringerTheses}, and can be obtained either by first-principles computation~\cite{Feibelman:1982,LiebschBook}, semiclassical models, or experiments~\cite{Yang:2019}.

The electromagnetic excitations of any system can be obtained by analyzing the poles of the (composite) system's scattering coefficients.
For the AGPs of a GDM structure, the relevant coefficient is the $p$-polarized reflection (or transmission) coefficient, whose poles are given by 
${ 1 \, - \, r_p^{2|\text{g}|1} \, r_p^{2|\text{m}} \, \e^{\iu 2 k_{z,2} t} = 0 }$~\cite{Chew_book}. 
Here, $r_p^{2|\text{g}|1}$ and $r_p^{2|\text{m}}$ denote the $p$-polarized reflection coefficients for the dielectric--graphene--dielectric and the dielectric--metal interface (detailed in Supplementary Note~2), respectively. Each coefficient yields a material-specific contribution to the overall quantum response: $r_p^{2|\text{g}|1}$ incorporates graphene's via $\sigma(q,\omega)$, and $r_p^{2|\text{m}}$ incorporates the metal's via the $d$-parameters (see Supplementary Note~2). 
The complex exponential [with $k_{z,2} \equiv (\ep_2 k_0^2 - q^2)^{1/2}$, where $q$ denotes the in-plane wavevector] incorporates the effects of multiple reflections within the slab. 
Thus, using the above-noted reflection coefficients (defined explicitly in the Supplementary Note~2), we obtain a quantum-corrected AGP dispersion equation:
\begin{widetext}
\begin{align}
 &\left[ \frac{\ep_1}{\kappa_1} + \frac{\ep_2}{\kappa_2} + \frac{\iu \sigma}{\omega \ep_0} \right] 
 \left[ \vphantom{\frac{\ep_j}{\kappa_j}}  \ep_\text{m} \kappa_2 + \ep_2 \kappa_\text{m} - \big( \ep_\text{m} - \ep_2 \big) \big(  q^2 d_\perp - \kappa_2 \kappa_\text{m} d_\pll \big) \right] = 
 \left[ \frac{\ep_1}{\kappa_1} - \frac{\ep_2}{\kappa_2} + \frac{\iu \sigma}{\omega \ep_0} \right] 
 \left[ \vphantom{\frac{\ep_j}{\kappa_j}}  \ep_\text{m} \kappa_2 - \ep_2 \kappa_\text{m} + \big( \ep_\text{m} - \ep_2 \big) \big(  q^2 d_\perp + \kappa_2 \kappa_\text{m} d_\pll \big) \right] \e^{-2 \kappa_2 t} 
  , 
  \label{eq:DispRel_DGDM_general}
\end{align}
\end{widetext}
for in-plane AGP wavevector $q$ and out-of-plane confinement factors $\kappa_j \equiv \sqrt{q^2 - \ep_j k_0^2}$ for $j \in \{1,2,\text{m}\}$. 

Since AGPs are exceptionally subwavelength (with confinement factors up to almost 300)~\cite{Lundeberg:2017,Iranzo:2018,Epstein:2020}, the nonretarded limit (wherein $\kappa_j \to q$) constitutes an excellent approximation.
In this regime, and for encapsulated graphene, \ie, where $\ep_\text{d} \equiv \ep_1 = \ep_2$, Eq.~\eqref{eq:DispRel_DGDM_general} simplifies to 
\begin{equation}
 \left[ 1 + \frac{2 \ep_\text{d}}{q} \frac{\omega \ep_0}{\iu\sigma} \right] 
 \left[ \vphantom{\frac{\ep_j}{\kappa_j}}  \frac{\ep_\text{m} + \ep_\text{d}}{\ep_\text{m} - \ep_\text{d}} -  q \big( d_\perp - d_\pll \big) \right] 
 =\Bigg[  1 +  q \big( d_\perp + d_\pll \big) \Bigg] \e^{-2 q t} 
 . \label{eq:DispRel_DGDM_symmetric_nonret}
\end{equation}

\begin{figure*}[t]
 \centering
  \includegraphics[width=0.75\textwidth]{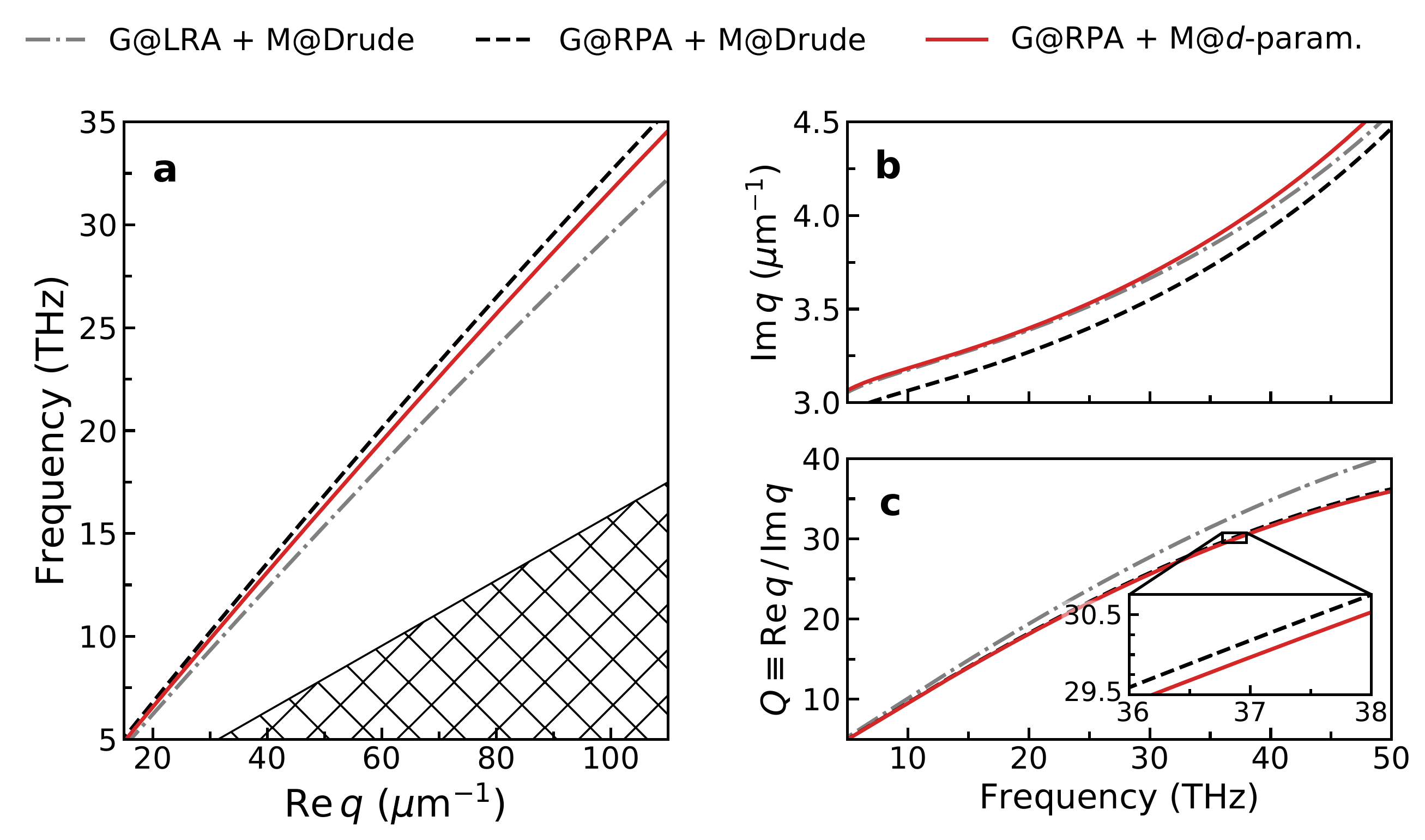}
  \caption{\textbf{Influence of metallic quantum surface-response on the dispersion of acoustic graphene plasmons (AGPs).} 
  Three increasingly sophisticated tiers of response models are considered: 
  (i)~classical, local response for both the graphene and the metal [gray dot-dashed line];
  (ii)~nonlocal RPA and local Drude response for the graphene and the metal, respectively [black dashed line]; and 
  (iii)~nonlocal RPA and $d$-parameter-augmented response for the graphene and the metal, respectively [red solid line].
  \textbf{a},~AGP dispersion diagram, $\omega/(2\pi)$ versus $\Re q$.
  The hatched region indicates the graphene's electron--hole continuum.
  \textbf{b},~Associated imaginary part of the AGP wavevector, $\Im q$.
  \textbf{c},~Corresponding quality factor $Q \equiv \Re q / \Im q$.
  The inset shows a zoom of the indicated region.
  System parameters: we take a graphene--metal separation of $t=\SI{1}{\nm}$; for concreteness and simplicity, we consider an unscreened jellium metal with plasma frequency $\hbar\omega_{\text{p}} \approx \SI{9.07}{\eV}$ (corresponding to $r_s=3$) where $\zeta \approx \SI{0.8}{\angstrom}$ and $\xi \approx \SI{0.3}{\angstrom}$~\cite{Persson:1983}, with Drude-type damping $\hbar\gamma_\text{m}=\SI{0.1}{\eV}$; for graphene, we take $E_\text{F}=\SI{0.3}{\eV}$ and $\hbar\gamma=\SI{8}{\meV}$; finally, we have assumed $\ep_\text{d}\equiv\ep_1=\ep_2=1$ (for consistency with the $d$-parameter data which assumes a metal--vacuum interface~\cite{Persson:1983}). 
  }\label{fig:GDM_Feib_dispersion}
\end{figure*}

For simplicity and concreteness, we will consider a simple jellium treatment of the metal such that $d_\parallel$ vanishes due to charge neutrality~\cite{Apell:1981,LiebschBook}, leaving only $d_\perp$ nonzero.
Next, we exploit the fact that AGPs typically span frequencies across the terahertz (THz) and mid-infrared (mid-IR) spectral ranges, \ie, well below the plasma frequency $\omega_{\text{p}}$ of the metal. 
In this low-frequency regime{, $\omega \ll \omega_\text{p}$}, the frequency-dependence of $d_\perp$ (and $d_\pll$) has the universal, asymptotic dependence 
\begin{equation}
 d_\perp (\omega) \simeq \zeta + \iu \frac{\omega}{\omega_\text{p}} \xi \  \qquad (\text{for } \ \omega \ll \omega_\text{p} ) 
 ,  \label{eq:d_perp_lowFreqs}
\end{equation}
as shown by Persson et al.~\cite{Persson:1983,Persson:1984} by exploiting Kramers--Kronig relations. 
Here, $\zeta$ is the so-called static image-plane position, \ie, the centroid of induced charge under a static, external field~\cite{Lang:1973}; and $\xi$ defines a phase-space coefficient for low-frequency electron--hole pair creation, whose rate is ${\propto}\,q\omega \xi$~\cite{LiebschBook}: both are ground-state quantities. 
In the jellium approximation of the interacting electron liquid, the constants $\zeta \equiv \zeta(r_s)$ and $\xi \equiv \xi(r_s)$ depend solely on the carrier density $n_{\text{e}}$, here parameterized by the Wigner--Seitz radius $r_sa_{\text{\textsc{b}}} \equiv ( 3 n_{\text{e}}/4 \pi )^{1/3}$ (Bohr radius, $a_{\text{\textsc{b}}}$). 
In the following, we exploit the simple asymptotic relation in Eq.~\eqref{eq:d_perp_lowFreqs} to calculate the dispersion of AGPs with metallic (in addition to graphene's) quantum response included.

\hfill

\noindent
\textbf{Quantum corrections in AGPs due to metallic quantum surface-response.} The spectrum of AGPs calculated classically and with quantum corrections is shown in Figure~\ref{fig:GDM_Feib_dispersion}.
Three models are considered: one, a completely classical, local-response approximation (LRA) treatment of both the graphene and the metal; and two others, in which graphene's response is treated by the nonlocal RPA~\cite{Wunsch:2006,Hwang:2007,Jablan:2009,GoncalvesPeres,Goncalves_SpringerTheses} while the metal's response is treated either classically or with quantum surface-response included (via the $d_\perp$-parameter). 
As noted previously, we adopt a jellium approximation for the $d_\perp$-parameter. 
Figure~\ref{fig:GDM_Feib_dispersion}a shows that---for a fixed wavevector---the AGP's resonance blueshifts upon inclusion of graphene's quantum response, followed by a redshift due to the quantum surface-response of the metal (since $\Re d_\perp > 0$ for jellium metals; electronic spill-out)~\cite{Liebsch:1987,Liebsch:1993,LiebschBook,Christensen:2017,Yang:2019,Goncalves:2020}.
This redshifting due to the metal's quantum surface-response is opposite to that predicted by the semiclassical hydrodynamic model (HDM) where the result is always a blueshift~\cite{Dias:2018} (corresponding to $\Re d_\perp^{\text{\textsc{hdm}}} < 0$; electronic ``spill-in'') due to the neglect of spill-out effects~\cite{Raza:2015a}. 
The imaginary part of the AGP's wavevector (that characterizes the mode's propagation length) is shown in Fig.~\ref{fig:GDM_Feib_dispersion}b: the net effect of the inclusion of $d_\perp$ is a small, albeit consistent, increase of this imaginary component. Notwithstanding this, the modification of $\Im q$ is not independent of the shift in $\Re q$; as a result, an increase in $\Im q$ does not necessarily imply the presence of a significant quantum decay channel [\eg, an increase of $\Im q$ can simply result from increased classical loss (\ie, arising from local-response alone) at the newly shifted $\Re q$ position]. 
Because of this, we inspect the quality factor $Q \equiv {\Re q}/{\Im q}$ (or ``inverse damping ratio''~\cite{Woessner:2015,Low:2017}) instead~\cite{Ni:2018} (Fig.~\ref{fig:GDM_Feib_dispersion}c), which provides a complementary perspective that emphasizes the effective (or normalized) propagation length rather than the absolute length. 
The incorporation of quantum mechanical effects, first in graphene alone, and then in both graphene and metal, reduces the AGP's quality factor.
Still, the impact of metal-related quantum losses in the latter is negligible, as evidenced by the nearly overlapping black and red curves in Fig.~\ref{fig:GDM_Feib_dispersion}c.

To better understand these observations, we treat the AGP's $q$-shift due to the metal's quantum surface-response as a perturbation: 
writing $q = q_{0} + q_{1} $, we find that the quantum correction from the metal is $q_{1} \simeq q_{0} d_\perp/(2 t)$,  
for a jellium adjacent to vacuum in the $\omega^2/\omega_{\text{p}}^2 \ll q_0t \ll 1$ limit (Supplementary Note~3). 
This simple result, together with Eq.~\eqref{eq:d_perp_lowFreqs}, provides a near-quantitative account of the AGP dispersion shifts due to metallic quantum surface-response: 
for $\omega \ll \omega_\text{p}$,  
\begin{enumerate*}[label=(\roman*)]
 \item $\Re d_\perp$ tends to a finite value, $\zeta$, which increases (decreases) $\Re q$ for $\zeta>0$ ($\zeta<0$); and 
 \item $\Im d_\perp$ is $\appropto\omega$ and therefore asymptotically vanishing as $\omega/\omega_{\text{p}}\rightarrow 0$ and so only negligibly increases $\Im q$.
\end{enumerate*} 
Moreover, the preceding perturbative analysis warrants ${\Re q_1}/{\Re q_0} \approx {\Im q_1}/{\Im q_0}$ (Supplementary Note~3), which elucidates the reason why the AGP's quality factor remains essentially unaffected by the inclusion of metallic quantum surface-response.
Notably, these results explain recent experimental observations that found appreciable spectral shifts but negligible additional broadening due to quantum response in the metallic substrate~\cite{Iranzo:2018,Epstein:2020}.

\begin{figure}[t]
 \centering
  \includegraphics[width=1.0\columnwidth]{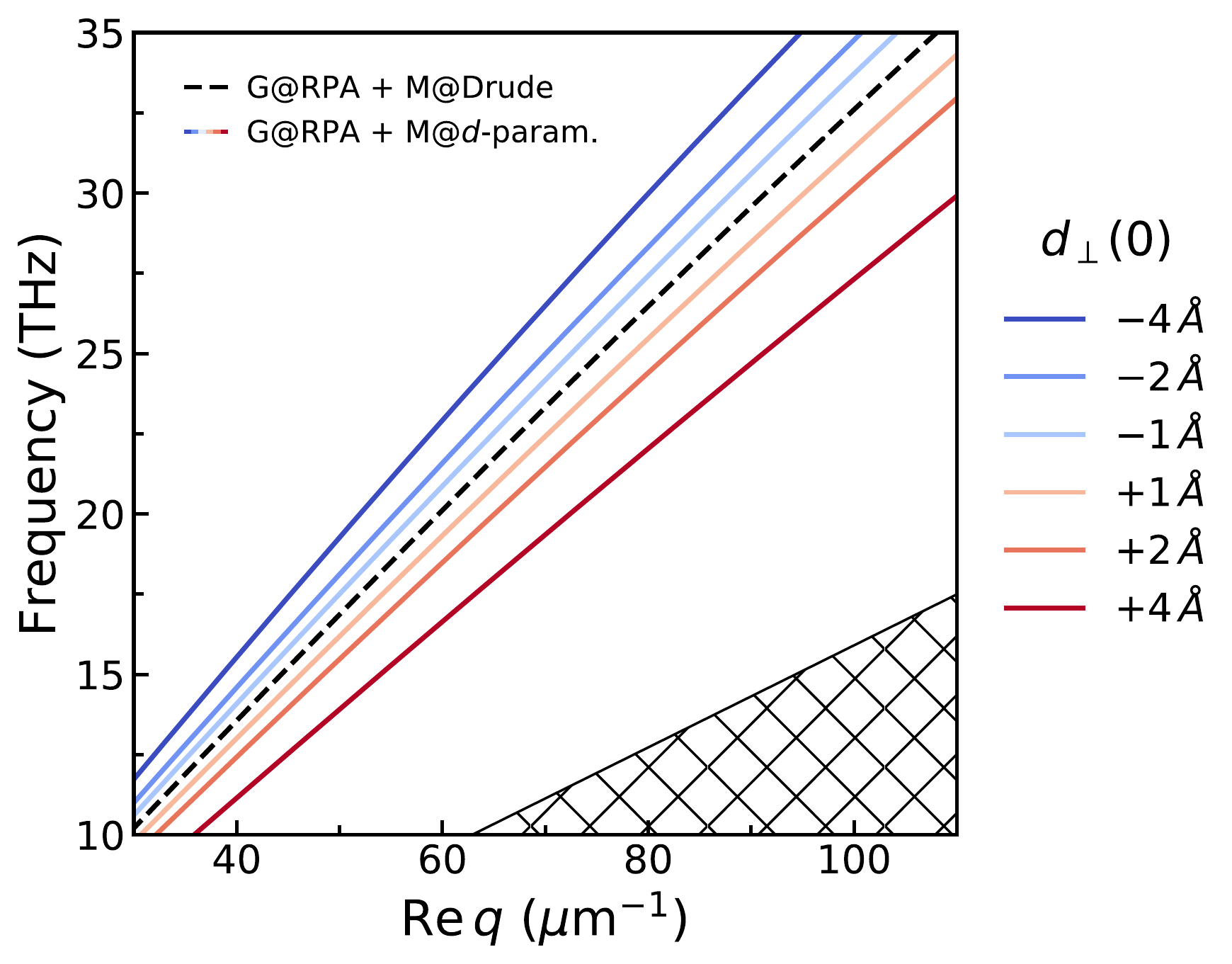}
  \caption{\textbf{Concept for using the spectral shifting of AGPs for retrieving the quantum surface-response of metals.}
  Impact of $d_\perp(\omega\ll\omega_{\mathrm{p}}) \simeq d_\perp(0) \equiv \zeta$ on the AGP's dispersion [obtained through the numerical solution of Eq.~(\ref{eq:DispRel_DGDM_general})].
All parameters (with the  exception of $d_\perp$) are the same as in Fig.~\ref{fig:GDM_Feib_dispersion}. 
  }\label{fig:GDM_Feib_varying_dperpStat}
\end{figure}

Next, by considering the separation between graphene and the metallic interface as a renormalizable parameter, we find a complementary and instructive perspective on the impact of metallic quantum surface-response. 
Specifically, within the spectral range of interest for AGPs (\ie, $\omega \ll \omega_{\text{p}}$), we find that the ``bare'' graphene--metal separation $t$ is effectively renormalized due to the metal's quantum surface-response from $t$ to $\tilde{t} \equiv t-s$, where $s\simeq d_{\perp}\simeq \zeta$ (see Supplementary Note~4), corresponding to a physical picture where the metal's interface lies at the centroid of its induced density (\ie, $\Re d_\perp$) rather than at its ``classical'' jellium edge. With this approach, the form of the dispersion equation is unchanged but references the renormalized separation $\tilde{t}$ instead of its bare counterpart $t$, \ie:
\begin{equation}
 1 + \frac{2 \ep_{\text{d}}}{q} \frac{\omega \ep_0}{\iu \sigma} = \frac{ \ep_{\text{m}} - \ep_{\text{d}} }{ \ep_{\text{m}} + \ep_{\text{d}} } \, \e^{-2 q \tilde{t}} 
 , \label{eq:delta_0}
\end{equation}
This perspective, for instance, has substantial implications for the analysis and understanding of plasmon rulers~\cite{Snnichsen:2005,Hill:2012,Teperik:2013} at nanometric scales.

Furthermore, our findings additionally suggest an interesting experimental opportunity: as all other experimental parameters can be well-characterized by independent means (including the nonlocal conductivity of graphene), high-precision measurements of the AGP's dispersion can enable the characterization of the low-frequency metallic quantum response---a regime that has otherwise been inaccessible in conventional metal-only plasmonics. 
The underlying idea is illustrated in Fig.~\ref{fig:GDM_Feib_varying_dperpStat}; depending on the sign of the static asymptote $\zeta \equiv d_\perp(0)$, the AGP's dispersion shifts toward larger $q$ (smaller $\omega$; redshift) for $\zeta > 0$ and toward smaller $q$ (larger $\omega$; blueshift) for $\zeta<0$.
As noted above, the $q$-shift is ${\sim}\, q_0 \zeta/(2t)$.
Crucially, despite the \r{a}ngstr\"{o}m-scale of $\zeta$, this shift can be sizable: 
the inverse scaling with the spacer thickness $t$ effectively amplifies the attainable shifts in $q$, reaching up to several \si{\per\micro\metre} for few-nanometer $t$.
We stress that these regimes are well within current state-of-the-art experimental capabilities~\cite{Lundeberg:2017,Iranzo:2018,Epstein:2020}, suggesting a new path toward the systematic exploration of the static quantum response of metals.

\begin{figure}[t]
 \centering
  \includegraphics[width=1.0\columnwidth]{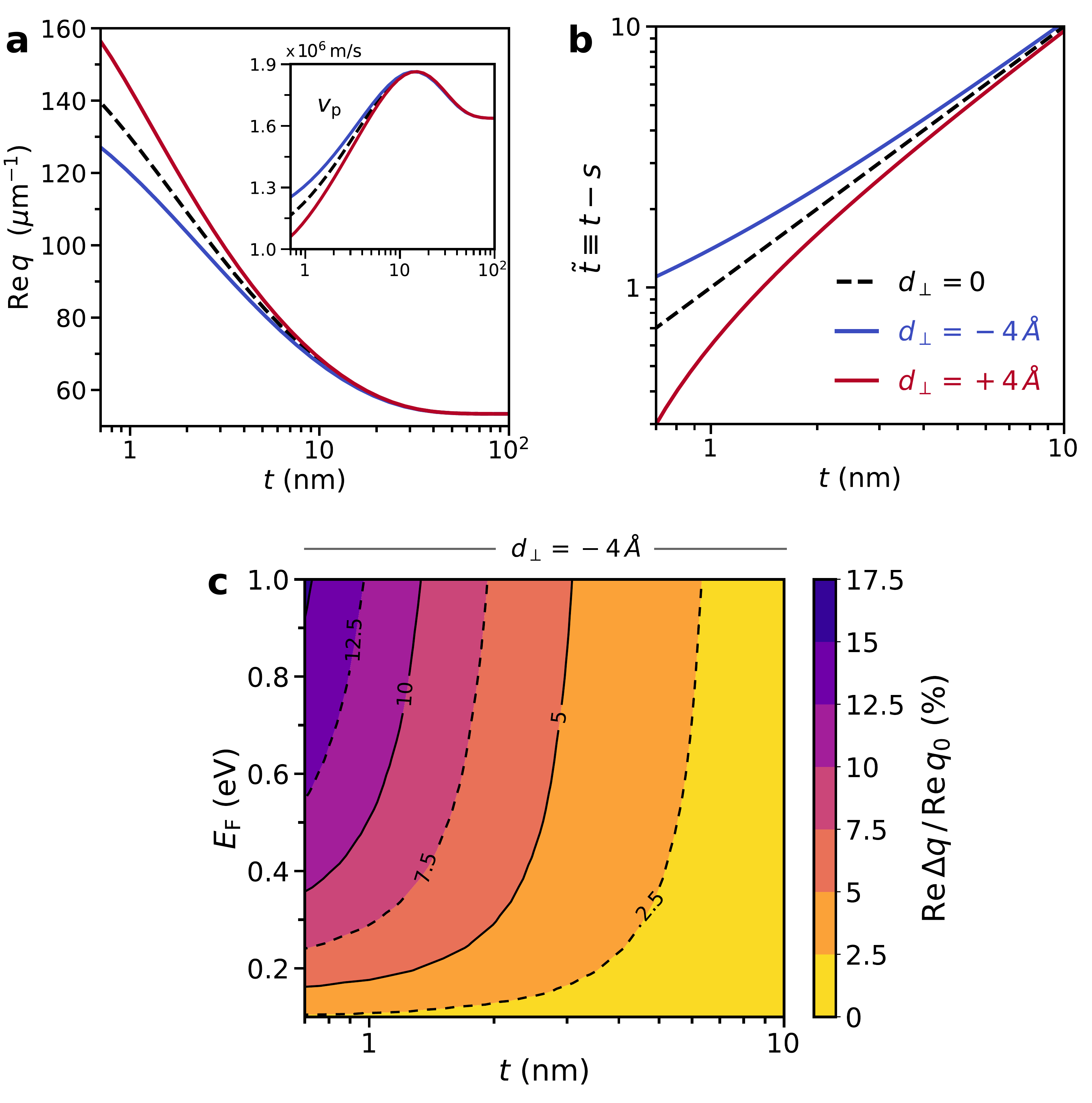}
  \caption{\textbf{Nonclassical corrections probed by AGPs.} 
  \textbf{a},~AGP's wavevector as a function of the graphene--metal separation $t$, contrasting the metal's response based on classical ($d_\perp=0$) and quantum ($d_\perp=\pm \SI{4}{\AA}$) treatments. Inset: corresponding group velocity $v_\text{p} = \partial \omega/\partial q |_{q=q(\omega_0)}$.
  \textbf{b},~Dependence of the renormalized graphene--metal separation $\tilde{t} \equiv t -s $ versus $t$.
  Setup parameters: $r_s=3$, $\hbar\gamma_{\text{m}}=\SI{0.1}{\eV}$, $\ep_{\text{d}}=4$, $\EF=\SI{0.3}{\eV}$ and $\hbar\gamma=\SI{8}{\meV}$; 
  we assume an excitation at $\lambda_0 = \SI{11.28}{\um}$  ($\hbar\omega_0 \approx \SI{110}{\meV}$ or $f_0 \approx \SI{26.6}{\THz}$)~\cite{Ni:2018}. 
  \textbf{c},~Relative quantum shift of the AGP wavevector, $\Re \Delta q /\Re q_0$, with $\Delta q \equiv q_0 - q$ where $q_0$ and $q$ denote the AGP wavevector associated with $d_\perp = 0$ and $d_\perp =-\SI{4}{\AA} $, respectively. The results presented in both \textbf{a} and \textbf{c} are based on the exact, numerical solution of Eq.~(\ref{eq:DispRel_DGDM_general}). 
  }
  \label{fig:GDM_Feib_varying_t}
\end{figure}

\hfill

\noindent
\textbf{Probing the quantum surface-response of metals with AGPs.} The key parameter that regulates the impact of quantum surface corrections stemming from the metal is the graphene--metal separation, $t$ (analogously to the observations of nonclassical effects in conventional plasmons at narrow metal gaps~\cite{Zhu:2016,Christensen:2017,Raza:2013}); see Fig.~\ref{fig:GDM_Feib_varying_t}.
For the experimentally-representative parameters indicated in Fig.~\ref{fig:GDM_Feib_varying_t}, these come into effect for $t \lesssim \SI{5}{\nm}$, growing rapidly upon decreasing the graphene--metal separation further.
Chiefly, ignoring the nonlocal response of the metal leads to a consistent overestimation (underestimation) of AGP's wavevector (group velocity) for $d_\perp < 0$, and vice-versa for $d_\perp > 0$ (Fig.~\ref{fig:GDM_Feib_varying_t}a); this behavior is consistent with the effective renormalization of the graphene--metal separation mentioned earlier (Fig.~\ref{fig:GDM_Feib_varying_t}b). 
Finally, we analyze the interplay of both $t$ and $\EF$ and their joint influence on the magnitude of the quantum corrections from the metal (we take $d_\perp = -\SI{4}{\angstrom}$, which is reasonable for the Au substrate used in recent AGP experiments~\cite{Alonso:2017,Lundeberg:2017,Iranzo:2018}); in Fig.~\ref{fig:GDM_Feib_varying_t}c we show the relative wavevector quantum shift (excited at $\lambda_0 = \SI{11.28}{\um}$~\cite{Ni:2018}).
In the few-nanometer regime, the quantum corrections to the AGP wavevector approach $5\%$, increasing further as $t$ decreases---for instance, in the extreme, one-atom-thick limit ($t \approx \SI{0.7}{\nm}$~\cite{Iranzo:2018}, which also approximately coincides with edge of the validity of the $d$-parameter framework, \ie, $t \gtrsim \SI{1}{\nm}$~\cite{Yang:2019}) the AGP's wavevector can change by as much as $10\%$ for moderate graphene doping. 
The pronounced Fermi level dependence exhibited in Fig~\ref{fig:GDM_Feib_varying_t}c also suggests a complementary approach for measuring the metal's quantum surface-response even if an experimental parameter is unknown (although, as previously noted, all relevant experimental parameters can in fact be characterized using currently available techniques~\cite{Lundeberg:2017, Iranzo:2018, Epstein:2020, Yang:2019}): 
such an unknown variable can be fitted at low $E_{\text{F}}$ using the ``classical'' theory (\ie, with $d_\perp = d_\parallel = 0$), since the impact of metallic quantum response is negligible in that regime.
A parameter-free assessment of the metal's quantum surface-response can then be carried out subsequently by increasing $E_{\text{F}}$ (and with it, the metal-induced quantum shift).
We emphasize that this can be accomplished in the same device by doping graphene using standard electrostatic gating~\cite{Lundeberg:2017, Iranzo:2018, Epstein:2020}.%

\section{Discussion}

In this Article, we have presented a theoretical account that establishes and quantifies the influence of the metal's quantum response for AGPs in hybrid graphene--dielectric--metal structures.
We have demonstrated that the nanoscale confinement of electromagnetic fields inherent to AGPs can be harnessed to determine the quantum surface-response of metals in the THz and mid-IR spectral ranges (which is typically inaccessible with traditional metal-based plasmonics). 
Additionally, our findings elucidate and contextualize recent experiments~\cite{Iranzo:2018,Epstein:2020} that have reported the observation of nonclassical spectral shifting of AGP's due to metallic quantum response but without a clear concomitant increase of damping, even for atomically thin graphene--metal separations.
Our results also demonstrate that the metal's quantum surface-response needs to be rigorously accounted for---\eg, using the framework developed here---when searching for signatures of many-body effects in the graphene electron liquid imprinted in the spectrum of AGP's in GDM systems~\cite{Lundeberg:2017}, since the metal's quantum-surface response can lead to qualitatively similar dispersion shifts, as shown here. 
In passing, we emphasize that our framework can be readily generalized to more complex graphene--metal hybrid structures either by semi-analytical approaches (\eg, the Fourier modal-method~\cite{NumnMethPhoton_book} for periodically nanopatterned systems) or by direct implementation in commercially available numerical solvers (see Refs.~\cite{Yang:2019,Echarri:2020}), simply by adopting $d$-parameter-corrected boundary conditions~\cite{Yang:2019,Goncalves:2020}.

Further, our formalism provides a transparent theoretical foundation for guiding experimental measurements of the quantum surface-response of metals using AGPs. 
The quantitative knowledge of the metal's low-frequency, static quantum response is of practical utility in a plethora of scenarios, enabling, for instance, the incorporation of leading-order quantum corrections to the classical electrostatic image theory of particle--surface interaction~\cite{Feibelman:1982} as well as to the van der Waals interaction~\cite{LiebschBook,Zaremba:1976,Persson:1984} affecting atoms or molecules near metal surfaces. 
Another prospect suggested by our findings is the experimental determination of $\zeta \equiv d_\perp(0)$ through measurements of the AGP's spectrum. This highlights a new metric for comparing the fidelity of first-principle calculations of different metals (inasmuch as \emph{ab initio} methods can yield disparate results depending on the chosen scheme or functional)~\cite{Mata:2017,Lejaeghere:2016} with explicit measurements.

Our results also highlight that AGPs can be extremely sensitive probes for nanometrology as plasmon rulers, while simultaneously underscoring the importance of incorporating quantum response in the characterization of such rulers at (sub)nanometric scales.
Finally, the theory introduced here further suggests additional directions for exploiting AGP's high-sensitivity, \eg, to explore the physics governing the complex electron dynamics at the surfaces of superconductors~\cite{Costa:2021} and other strongly-correlated systems.

\vfill


\newpage

%
\def\bibsection{\section*{\refname}} 

%

\newpage


\hfill

\noindent\large\textbf{Acknowledgements}
{\small 
\begin{acknowledgments}
\noindent
N.\,A.\,M. is a VILLUM Investigator supported by VILLUM FONDEN (Grant No.~16498) and Independent Research Fund Denmark (Grant No.~7026-00117B).
The Center for Nano Optics is financially supported by the University of Southern Denmark (SDU~2020 funding). 
The Center for Nanostructured Graphene (CNG) is sponsored by the Danish National Research Foundation (Project No.~DNRF103). 
This work was partly supported by the Army Research Office through the Institute for Soldier Nanotechnologies under contract No.~W911NF-18-2-0048. 
N.\,M.\,R.\,P. acknowledges support from the European Commission through the project ``Graphene-Driven Revolutions in ICT and Beyond'' (No.~881603, Core~3), COMPETE~2020, PORTUGAL~2020, FEDER and the Portuguese Foundation for Science and Technology (FCT) through project POCI-01-0145-FEDER-028114 and through the framework of the Strategic Financing UID/FIS/04650/2019. 
F.\,H.\,L.\,K. acknowledges financial support from the Government of Catalonia trough the SGR grant and from the Spanish Ministry of Economy and Competitiveness (MINECO) through the Severo Ochoa Programme for Centres of Excellence in R\&D (SEV-2015-0522), support by Fundaci\'{o} Cellex Barcelona, Generalitat de Catalunya through the CERCA program,  and the MINECO grants Plan Nacional (FIS2016-81044-P) and the Agency for Management of University and Research Grants (AGAUR) 2017 SGR 1656.  Furthermore, the research leading to these results has received funding from the European Union’s Horizon 2020 program under the Graphene Flagship grant agreements No.~785219 (Core~2) and No.~881603 (Core~3), and the Quantum Flagship grant No.~820378. This work was also supported by the ERC TOPONANOP (Grant No.~726001).
\end{acknowledgments}
}

\end{document}